# Weak magnetic order in the normal state of the high-$T_c$ superconductor La$_{2-x}$Sr$_x$CuO$_4$

C. Panagopoulos[1,2], M. Majoros[2], T. Nishizaki[3], and H. Iwasaki[4]

[1] *Cavendish Laboratory, University of Cambridge, Cambridge CB3 0HE, UK*
[2] *IRC in Superconductivity, University of Cambridge, Cambridge CB3 0HE, UK*
[3] *Institute of Materials Research, Tohoku University, Sendai 980-8577, Japan*
[4] *School of Materials Science, JAIST, Tatsunokuchi 923-12, Japan*

We report magnetization measurements in the normal state of the high transition temperature (high-$T_c$) superconductor La$_{2-x}$Sr$_x$CuO$_4$. A magnetic order in the form of hysteresis in the low-field magnetization is observed at temperatures well above $T_c$. The doping ($x$) dependence of the onset and strength of this order follows $T_c(x)$ and falls within the pseudo-gap regime.

PACS numbers: 74.25.Dw, 74.72.Dn, 74.25.Ha, 74.40.+k

*I. Introduction* – Resolving the high-$T_c$ mechanism depends primarily on understanding: (i) The evolution of the doped Mott insulator with doping ($x$), and (ii) The origin of the pseudogap, characterized by partial gapping of the low energy density of states, below a characteristic temperature $T^*(x) > T_c(x)$ [1-6]. Proposed scenarios for the pseudogap state include static or fluctuating ordering phenomena, such as charge, spin or orbital current order, or local pairing correlations with a phase which is locally well defined, supporting the presence of superconducting domains/vortices [2,3]. Transport measurements on YBa$_2$Cu$_3$O$_{7-x}$ nanowires suggested the presence of some form of domains, reflected as cooperative telegraph-like fluctuations and thermal hysteresis at $T<T^*$ [7]. Similarly, experiments on bulk La$_{2-x}$Sr$_x$CuO$_4$ crystals revealed the presence of a hysteresis in the temperature dependence of the low-field magnetization [8,9]. Moreover, the thermal hysteresis was found to be associated with superconductivity-related long-lived persistent currents up to a characteristic temperature $T_c<T_s≤T^*$ [9]. These results are suggestive of some form of superconductivity-related charge and spin order, present in the normal state of the high temperature superconductors (HTS).

In this Letter we report detailed magnetization measurements, performed using a superconducting quantum interference device (SQUID) with sensitivity better than $10^{-8}$ emu on a series of extensively characterized [8-17], both single and polycrystalline La$_{2-x}$Sr$_x$CuO$_4$ samples. We have identified for the first time the presence of a well-developed magnetic order in the pseudogap phase. The observed order is associated with the aforementioned hysteretic effects, and shows systematic trends with doping resembling the superconducting dome of HTS.

*II. Experimental* - The preparation and characterization of the polycrystalline samples ($x$=0.03-0.24) has been discussed in [10-15], and for the single crystals ($x$=0.03, 0.15) in [16,17]. Chemical and elemental analysis showed them to be phase pure and stoichiometric. Over the years, the high purity of the polycrystalline samples and the absence of any phase that could possibly contribute to our observations have been confirmed by several transport, thermodynamic and spectroscopic measurements [11-15]. The single crystals have also been characterized and studied by spectroscopic and thermodynamic methods [16,17]. The sample high quality and the intrinsic, bulk nature of the hysteresis have been further tested and discussed in [8-11]. Magnetization measurements were performed using a MPMS-*XL* SQUID magnetometer. As discussed in [8,9] the combination of low noise, absence of background, controlled slow temperature and field sweeps, and high statistics gave a resolution $10^{-8}$ emu or better. The hystereses reported here have been measured using both the DC and the Reciprocating Sample Options, the latter suitable for improving the signal/noise ratio allowing accurate estimates for the size of the hysteresis.

*III. Doping dependent magnetic hysteresis* – Customarily, the doping dependence of the normal state magnetization of HTS is investigated by cooling the sample in a fixed high field (several Tesla) and measuring its temperature dependence, or studying the field dependence in high magnetic fields [1-6,18]. Our detailed low-field data has revealed a doping dependent deviation, in the form of a magnetic hysteresis, of the bulk magnetization $M$ from the high-field linearity (Fig. 1). The doping dependence of the sample geometry independent quantities coercive field, and normalized hysteresis width at the origin ($\Delta M/M_c$), where $M_c$ is the value of magnetization the hysteresis closes, show a correlation with $T_c(x)$ (the doping dependence of $T_c$ is shown in Fig. 3). We note the marked suppression in the $x=1/8$ region, just like in $T_c$ and the superfluid density [14]. The magnetic hysteresis is not unique to concentrations displaying bulk superconductivity. It is also observed for compositions lying near the onset of the superconducting dome, as shown in Fig. 1 for single crystal and polycrystalline samples with $x=0.03$ and $0.05$, respectively, at temperatures well above the respective spin glass temperatures [14].

The hysteresis loops in Fig. 1 are typical of magnetic domains *i.e.*, local moments which reverse with field polarity, and are distributed throughout the bulk of a sample [19]. In Fig. 2(a) we depict, for the polycrystalline $x$=0.10 sample, typical results for a system with domains [19]. In a material which consists of a number of domains, as $H$ is increased while on the virgin curve domains orient in the direction of the applied field [19,20]. With reducing $H$ a subset of these domains is no longer oriented, and subsequent increase in $H$ to the original value restores the original set of non-oriented domains demonstrating the return point memory effect [19-20]. It is the domains displaying return point memory which are responsible for the magnetic order we observe at $T>T_c$.

According to Preisach [20], if a system contains independent elementary hysteresis domains, two subloops taken while on the virgin curve and between the same applied field end points, must be congruent [19,20]. The presence of subloops is depicted in the main panel of Fig. 2(a) for $x$=0.10 (sample grown in Cambridge). At $T$=45K ($>T_c$) the subloops A and B shown in the lower inset are rotated relative to one another indicating the domains interact. Figure 2(b) shows another example of rotated subloops and return point memory, but now for a single crystal (grown in JAIST) with $x$=0.15 at $T$=70K ($>T_c$). Interestingly, when $T \geq 300K$ and $T \geq 270K$, for $x$=0.10 and 0.15 respectively, the subloops are congruent [Fig. 2(a) (upper inset) and Fig. 2(b) (inset)] and the hysteresis is suppressed.

Figure 3 shows the doping dependence of the temperature $T_{onset}$ below which a magnetic hysteresis develops. $T_{onset}(x)$ resembles $T_c(x)$, $\Delta M/M_c(x)$ (Fig. 1) and $T_s(x)$ - the onset of the thermal hysteresis observed recently on the same samples, and in the same temperature regime for as long the field is applied at $T < T_s \sim T_{onset}$ [8,9]. Furthermore, the observed magnetic state falls within the pseudogap phase. In fact the values of $T_{onset}(x$ $\geq 0.10$) are in good agreement with $T^*(x \geq 0.10)$ obtained by experiments probing the single particle excitation spectrum (to the best of our knowledge there is no data available for $T^*(0.03 \leq x < 0.10)$) [2,3]. Our data provide the first experimental evidence for an actual bulk order within the pseudogap regime of HTS, and with a doping dependence following $T_c$ broadly as expected for phase fluctuations encouraging the onset of bulk superconductivity [2,3].

*IV. Origin of the magnetic order* – The observed hysteresis is clearly due to a mechanism incorporating domains, which may be in the form of droplets (*e.g.* vortices) or rivers (*e.g.* stripes – constituting antiphase domain walls in the antiferromagnet) [1-3,19]. It is important we establish whether this order is at all associated with superconductivity, apart from its similar doping dependence to $T_c$. To this aim it is necessary to identify the currents responsible for the magnetic hysteresis, and the associated extension of a thermal hysteresis reported to temperatures well above the irreversibility temperature, and as high as $T_s(x) \sim T_{onset}(x)$ [8,9].

Experiments towards this aim have revealed remarkable similarities in the thermomagnetic behavior of $M$ below and above $T_c$ [8,9]. One such similarity was observed in measurements of the magnetization as a function of temperature by reducing the applied field to zero while in the mixed state: As expected, induced currents kept the superconducting vortices trapped inside the sample giving rise to a paramagnetic moment below $T_c$ [9]. However, although these currents decreased with increasing temperature, instead of vanishing near $T_c$ the moment survived up to $T_{onset}(x)$ indicating persistent currents, and suggesting trapped vortices up to these high temperatures [9]. Compelling evidence for vortices and diamagnetic fluctuations at $T >> T_c$ has in fact been reported [21,22] and shown to develop with doping just like the magnetic order does, but at somewhat lower temperatures (for $La_{2-x}Sr_xCuO_4$ up to $\sim 4T_c$). We note the high fields employed in those studies [21,22] may have suppressed the upper temperature limit of the onset of diamagnetic fluctuations.

Is there theoretical support for vortices at $T>T_c$? In systems where the pairing of electrons originates from strong repulsive interactions, local superconductivity allows for phase fluctuations to commence prior to bulk superconductivity, and persist over a wide temperature region at $T>T_c$ [2]. In the fluctuation regime there is short range phase coherence, and vortices may exist [2,3]. Although it is unclear whether phase fluctuations can extend well above $2T_c$, fluctuations surviving to $T>2T_c$ and vanishing gradually with increasing temperature [23], may in fact be consistent with the persistence of a Nernst signal and a diamagnetic response up to $4T_c$ [21,22]; with only a few vortices pinned by quenched disorder, giving rise to a hysteresis only at low fields (Fig.1) and surviving to higher temperatures. It may be that the combination of lower phase stiffness and less efficient screening in these electronically disordered systems might as well allow the presence of superconducting fluctuations to these high temperatures.

Based on the above discussion, $T_{onset}(x)$ would signify the onset of local superconductivity. If however we consider superconducting fluctuations, which are necessary in order to support the presence of superconducting vortices at these high temperatures, may extend strictly only up to $\sim 2T_c$, the present work (Figs 1-3) provides at least direct experimental evidence for a domain-driven magnetic order which develops with carrier concentration together with superconductivity, but at a higher temperature and within the pseudogap regime. Suggestions for such order have included heterogeneous charge, spin ordering in the presence of quenched disorder [2,23-26]. That the identified magnetism (Figs 1-3) is associated with persistent currents [9], develops like the Nernst signal [21], the diamagnetic response above $T_c$ [22], and the superconducting fluctuations [2,3,23], consistently suggests this order is most likely to be linked to and encourage, local and eventually global superconductivity at lower temperatures.

*V. Summary* – Through detailed sensitive low-field magnetization measurements, we have observed for the first time a well developed bulk magnetic order in the normal state of the high-$T_c$ superconductor $La_{2-x}Sr_xCuO_4$. The identified order falls within the pseudogap regime, and tracks the dome-shaped doping dependent superconducting transition temperature.

We thank E. Carlson, K. Dahmen, E. Fradkin, S. Kivelson and N. Papanicolaou for helpful discussions, and C. Bowell and A. Petrović for measurements at the early stages of this work. We have crosschecked our results on samples kindly lend by J. Cooper and T. Sasagawa. M.M. acknowledges the AFRL/PRPS Wright-Patterson Air Force Base, Ohio, for financial support. C.P. and the work in Cambridge were supported by The Royal Society.

FIG. 1 (Color online). Magnetization ($M$) as a function of field ($H$) for various doping ($x$) levels for $La_{2-x}Sr_xCuO_4$. The arrows indicate the trajectories followed in changing the applied field. Data for $x$=0.03 and 0.15 are for single crystals with $H//ab$. All other hysteresis data are for polycrystalline samples. The insets for $x$=0.07, 0.10 and 0.17 depict data at higher fields. The plots for $x$=0.05, 0.07, 0.10 and 0.15 include characteristic data for the virgin curves (in red). The lowest panel depicts the doping dependence of the coercive field (circles) and the normalized hysteresis width at the origin (squares).

FIG. 2 (Color online). Magnetic loops performed on the virgin curve for $La_{1.9}Sr_{0.10}CuO_4$ (polycrystal) and $La_{1.9}Sr_{0.15}CuO_4$ (single crystal). (a) The main plot shows a set of subloops for $x$=0.10 performed on the virgin curve at $T$=45K. The arrows denote the trajectories followed around the various loops. The lower inset compares the two subloops (A and B) shown on the main panel which were taken between the same applied field end points at $T$=45K. The upper inset shows subloops obtained the same way as in the lower inset but at $T$=300K. The subloops in the insets have been shifted to zero for comparison. (b) Low-field $M(H)$ for an $x$=0.15 single crystal ($T_c$=37.5K, $m$=1.1mg). Main panel: comparison of subloops performed on the virgin curve at $T$=70K the same way as for $x$=0.10. The trajectories followed, as well as the colors and symbols are in accordance with panel (a). The inset depicts the subloops performed on the virgin curve at $T$=275K.

FIG. 3 (Color online). Doping dependence of the onset temperature $T_{\text{onset}}$ (blue circles) of the normal state magnetic order and the superconducting transition temperature $T_c$ (black line) in $La_{2-x}Sr_xCuO_4$. Data for $x=0.03$ and $0.15$ are for single crystals with $H//ab$.

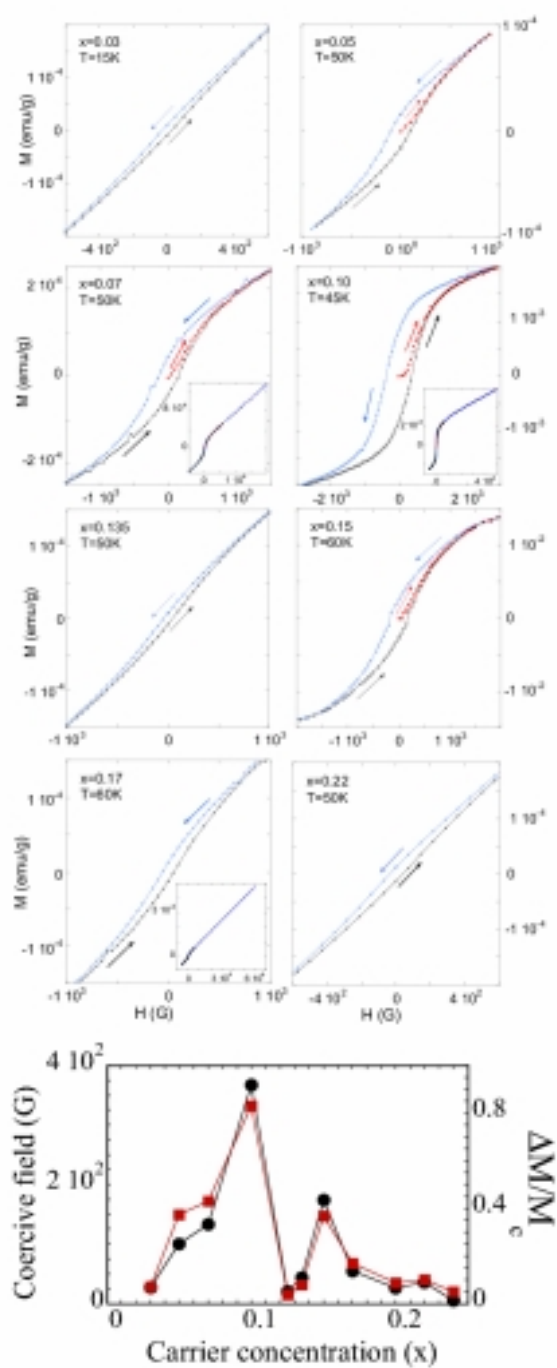

Panagopoulos_Fig. 1 Revised

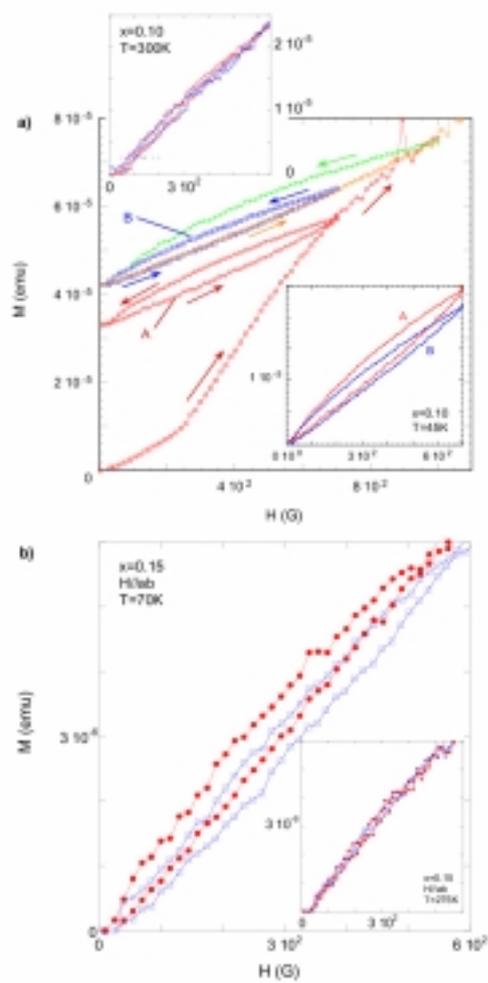

Panagopoulos_Fig. 2 Revised

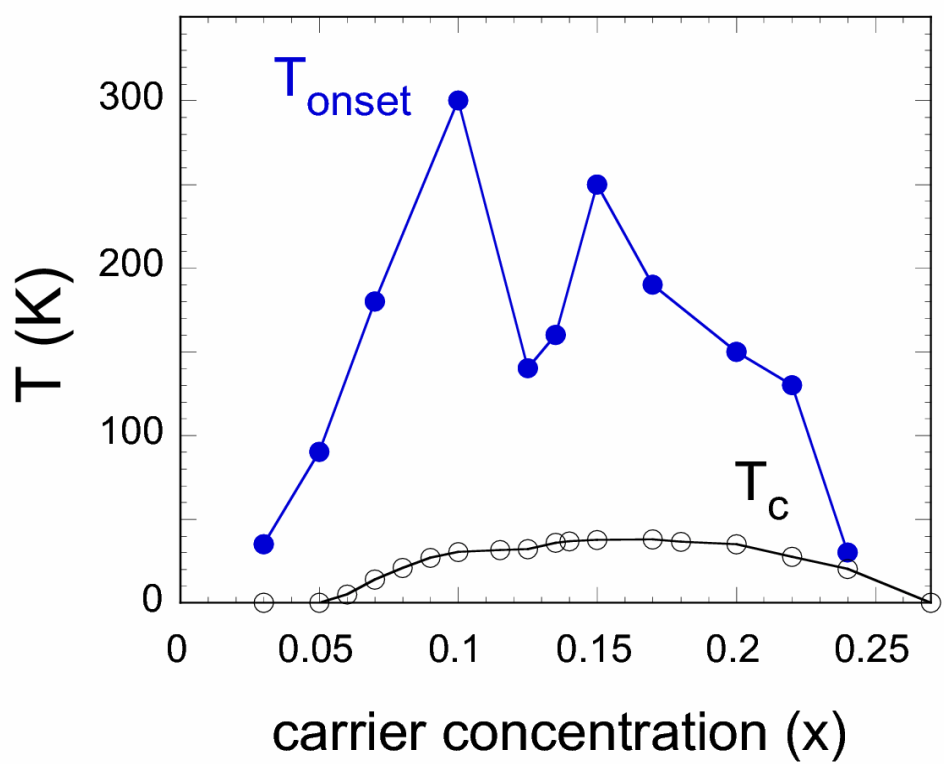

Panagopoulos_Fig. 3 Revised